\documentclass[conference]{IEEEtran}
\IEEEoverridecommandlockouts

\def\BibTeX{{\rm B\kern-.05em{\sc i\kern-.025em b}\kern-.08em
		T\kern-.1667em\lower.7ex\hbox{E}\kern-.125emX}}

\usepackage{subfig}
\usepackage{amsmath,amsfonts}
\usepackage{lscape}
\usepackage{multicol}
\usepackage{multirow}
\usepackage{booktabs}
\usepackage[colorlinks=true, allcolors=blue]{hyperref}
\usepackage{cite}
\usepackage{amsmath,amssymb,amsfonts}
\usepackage{algorithmic}
\usepackage{graphicx}
\usepackage{textcomp}
\usepackage{xcolor}

\begin{document}

	\title{DeepOrder: Deep Learning for Test Case Prioritization in Continuous Integration Testing}

	\author{\IEEEauthorblockN{Aizaz Sharif}
		\IEEEauthorblockA{\textit{Simula Research Laboratory}\\
			Norway \\
			aizaz@simula.no}
		\and
		\IEEEauthorblockN{Dusica Marijan}
		\IEEEauthorblockA{\textit{Simula Research Laboratory} \\
			Norway \\
			dusica@simula.no}
		\and
		\IEEEauthorblockN{Marius Liaaen}
		\IEEEauthorblockA{\textit{Cisco Systems} \\
			Norway }
	}

	\maketitle

	\begin{abstract}
		Continuous integration testing is an important step in the modern software engineering life cycle. Test prioritization is a method that can improve the efficiency of continuous integration testing by selecting test cases that can detect faults in the early stage of each cycle. As continuous integration testing produces voluminous test execution data, test history is a commonly used artifact in test prioritization. However, existing test prioritization techniques for continuous integration either cannot handle large test history or are optimized for using a limited number of historical test cycles. We show that such a limitation can decrease fault detection effectiveness of prioritized test suites.
		
		This work introduces DeepOrder, a deep learning-based model that works on the basis of regression machine learning. DeepOrder ranks test cases based on the historical record of test executions from any number of previous test cycles. DeepOrder learns failed test cases based on multiple factors including the duration and execution status of test cases. We experimentally show that deep neural networks, as a simple regression model, can be efficiently used for test case prioritization in continuous integration testing. DeepOrder is evaluated with respect to time-effectiveness and fault detection effectiveness in comparison with an industry practice and the state of the art approaches. The results show that DeepOrder outperforms the industry practice and state-of-the-art test prioritization approaches in terms of these two metrics.
	\end{abstract}


	\maketitle
	
	\section{Introduction}~\label{sec:Introduction}
	Continuous integration (CI)~\cite{R11} is a common and widely used software engineering development practice.  
	Each CI cycle involves software testing which aims to detect potential bugs in the changed code before deploying it to production. One of the important tasks performed in CI testing is regression testing where new code changes are tested within each CI cycle. 
	Relevant test cases have to be chosen from the entire test suite in order to detect potential bugs introduced by the new changes. However, in the industrial case study investigated in this paper, not all relevant tests can be run, even on the safe side, due to time constraints 
	inherent to CI. To address this challenge, Regression Test Prioritization (RTP) in CI suggests to rank test cases with the highest probability of detecting faults, which, in addition, should be able to detect faults as early as possible.
	
	\textit{CI Testing in Practice}: Our work is motivated by the practical problem of regression testing in CI at Cisco Systems Norway. Test engineers have a daily task to regression-test the latest version of video conferencing software, before the software can be integrated to the mainline codebase. Test engineers have available a set of several hundred test cases, a test execution log containing historical test failure information from previous CI cycles and an x-hour time frame to complete the task of selecting relevant most likely failing test cases for the next CI run (time frame changes from one to another CI cycle). Test cases are system-level, and testing does not require source
	code. This task imposes two challenges on test engineers. First, selecting test cases able to detect as many regression faults as possible. Second, fitting test selection and execution to a specific duration. 
	We have previously developed an automatic history-based test case prioritization for CI at Cisco, namely the ROCKET method \cite{R5}. However, as the amount of historical test data increased over time, ROCKET faced scalability issues, in terms of long runtime, as its main limitation.

	An extensive amount of research is done in the area of RTP in order to improve the effectiveness of 
	test prioritization. However, 
	the state-of-the-art (SoA) test case prioritization algorithms are optimized for using only limited test information~\cite{R3}, and often are not fast enough to perform in the industrial CI settings with a large number of test executions. 
	An appropriate algorithm for the RTP task in practice where voluminous test data is generated would be the one that not only performs failure-inducing test case prioritization while considering time constraints, but is also computationally inexpensive (this may have different interpretations in different industrial settings). 
	RETECS~\cite{R3} is a recent work addressing the RTP problem 
	in CI using 
	reinforcement learning. Even though RETECS achieved good results, the algorithm has two shortcomings. 
	First, it requires a significant amount of time for model learning. Second, it is optimized for using historical test data of only the last four test cycles, which turns out to decrease the fault detection effectiveness of prioritized test cases, as we show in our experiments.

	In this paper, we are proposing DeepOrder, a time-efficient deep-learning-based regression model for RTP  
	in CI. Using deep neural networks (DNN), we enable the use of test history from any number of previous CI cycles, thus addressing the limitation of the existing approaches for RTP in CI ~\cite{R3} ~\cite{R5}. 
	Compared to other available approaches, our approach (i) achieves higher fault detection effectiveness of prioritized test suites in a given time frame (compared to \cite{R3}), (ii) has lower training and prioritization time (compared to \cite{R3}), and (iii) achieves higher time-effectiveness of prioritized test suites and lower runtime when voluminous test history is used in RTP (compared to \cite{R5}).] 
	The contributions of this paper are:
	\begin{enumerate}
		\item Using supervised learning, in its simplest form, we use historical test data to train a regression-based deep learning model for RTP, which improves the industry practice of CI testing. 
		We further show that our model performs better than the SoA models.
		
		\item We show that using test history older than the last four cycles for RTP increases the fault detection effectiveness of prioritized test suites. History length in the current SoA machine learning based approaches for RTP in CI is only four cycles (usually because longer histories are expensive to treat). 
		
		\item 
		We experiment with the model on four industrial datasets, 
		showing the potential of our approach to be integrated into the CI regression testing in practice.

	\end{enumerate}

	\section{Problem Definition}~\label{sec:Formal}
	Our work is aimed at RTP in CI testing using historical test data. Influenced by our industrial case study, we deal with system-level test cases. The information of each test case is independent of the code for which the test case was written, 
	and therefore we ignore any information that is related to version control and codebase.
	
	\subsection{Formal Notations}~\label{sec:Objective}
	Given a test suite $T_S=\{t_1,\ldots,t_n\}$ as a set of test cases, their historical test case execution effectiveness (pass/fail status) $ES$, the set of possible test case orderings $P$ of $T_S$, and the objective function $f: P \rightarrow \mathbb{R}$, the RTP problem aims at finding $T' \in P$ such that 
	
	$$(\forall T'') (T'' \in P) (T'' \neq T') [f(T') \geq f(T'')]$$
	
	\noindent which means that the test case ordering $T'$ will achieve higher fault detection effectiveness in execution than the test case ordering $T''$. The priority of a test case $p(t_i)$ corresponds to its relative order within $T'$.  
	
	In addition to solving the RTP problem, there is often a need in practice (as it was the case in our industrial CI case study) for solving a time-aware test case selection (TA-TCS) problem. \textit{TA-TCS} can be defined as follows. Given a test suite $T_S$, a time budget $T_c$ (total available time for testing) and the average test execution duration of a test case $d(t_i)$, the goal is to select test cases $T\subseteq T_S$ such to $max(|T|)$ while
	
	$$\sum_{t_i \in T} d(t_i) \leq T_c$$ 
	
	\noindent which means to select as many test cases as possible such that their execution time does not exceed the time budget.

	Furthermore, we assume 
	that a test case is executed only once in a single CI cycle.
	Therefore, in a given CI cycle $j$, the {\it ExecutionStatus} $ES$ of a test case $t_i$ is 
	\[
	ES_{(i,j)} = 
	\begin{cases}
	1,& \text{if } t_i \text{ failed in cycle } j 
	\\
	0,& \text{if } t_i \text{ passed in cycle }  j 
	\\
	-1 & \text{if } t_i \text{ was not executed in cycle } j 
	\end{cases}
	\]
	
	Historical test case execution data can be represented as shown in Table \ref{CICyclePlot}.
	\begin{table}
		\begin{center}
			\small
			\caption{Historical test case execution statuses across $m$ CI cycles. Test case duration $d(t_i)$ is an average execution time across $m$ CI cycles. Test case priority $p({t_i})$ is calculated for the cycle $m+1$.}
			\label{CICyclePlot}
			\begin{tabular}{c||c|c|c||c||c}
				& $ES_{(i,1)}$  & ... & $ES_{(i,m)}$ & $d [sec]$  & $p$ \\
				\hline
				$t_1$ & 0      & ... & -1 & 105  & 0.5   \\
				$t_2$ & -1    & ... & 1  & 49  & 0.9   \\
				... & ... & ... & ... & ...  & ...   \\
				$t_n$ & 1      & ... & 0 & 96    & 0.3   \\
			\end{tabular}
		\end{center}
	\end{table}
	We have available test case execution statuses for the previous $m$ CI cycles. The $p$ value is the calculated test case priority in the cycle to-be-executed $m+1$. The $d$ value represents the average test case execution time across $m$ CI cycles, i.e. 
	$$\sum_{j \in 1..m} \frac{d(t_{(i,j)})}{m}$$

	\begin{figure*}[!bthp]
		\centering
		\includegraphics[width=5.3in, keepaspectratio]{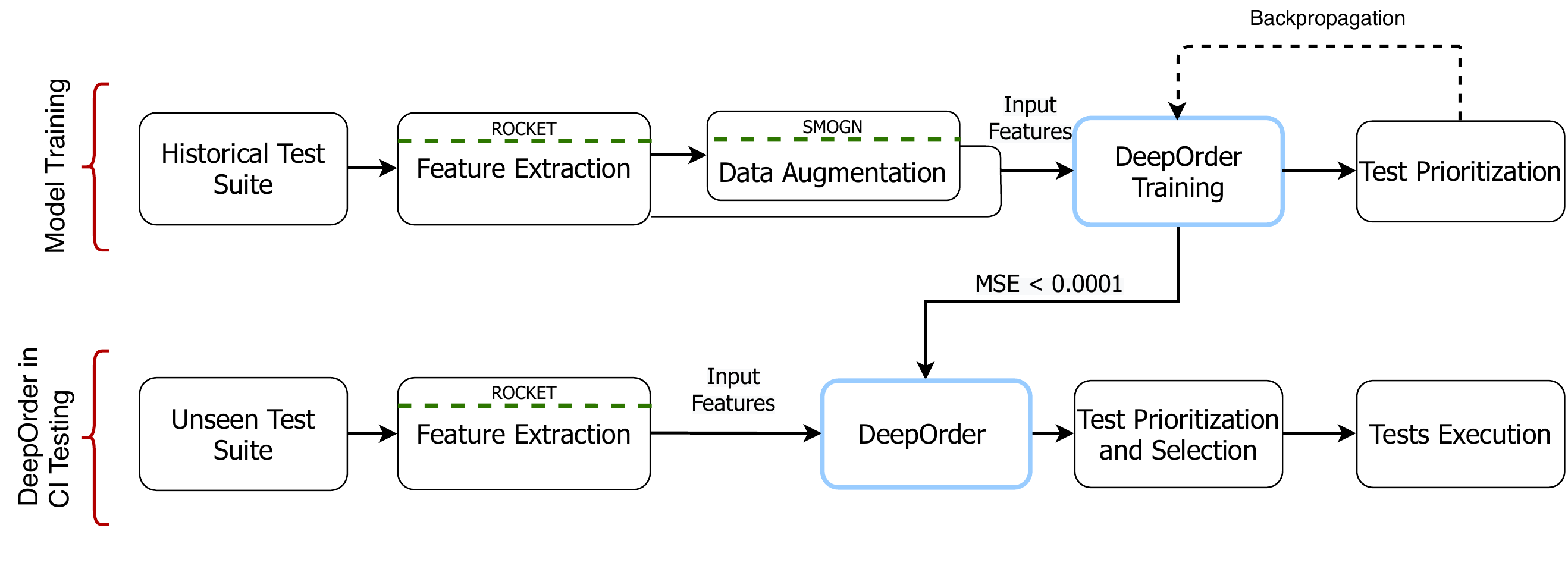}
		\caption{DeepOrder integrated in CI testing}
		\label{CICycle}
	\end{figure*} 
	
	\subsection{Test Prioritization in CI}~\label{sec:ROCKET}
	Our approach to test prioritization in CI consists in computing a test case priority using historical test case execution statuses and a test case execution duration. For each test case, given the status of its previous executions in successive CI cycles and its average execution time, the algorithm computes a priority value which 
	optimizes the following bi-objective function:
	$$g(maximize(f), maximize(|T|)) $$
	The first objective is to maximize the fault detection ability of a test suite, guided by the historical fault detection capability of test cases. The second objective is to maximize the number of executed test cases within a given time budget. These two objectives originate from the requirements of our industrial case study.
	
	In order to reflect the higher impact of more recent failures in CI cycles, a test case priority value $p(t_i)$ is computed as:
	$$p(t_i) = \sum_{j \in 1..m} \omega_j \times \max(ES_{(i,j)}, 0) $$
	where $\omega_j$ is an assigned weight (a real value in $(0,1)$) to each cycle $j$ such that $\sum_{j} \omega_j = 1$, and where recent test failures have a greater weight, and $m$ is the number of historical CI cycles. $p(t_i) \in (0,1)$. $\max(ES_{(i,j)},0)$ is used to discard the calculations for non-executed test cases in a given cycle (i.e., where $ES_{(i,j)} = -1$).

	\vspace{30pt}
	
	\section{DeepOrder: Using Deep Learning for Test Case Prioritization}~\label{sec:DeepOrder}
	\subsection{DeepOrder Approach and its Integration in CI testing}\label{sec:Overview} 
	DeepOrder is based on supervised deep learning and it exploits a trained multi-layer neural network model to learn an approximation function for prioritizing test cases. Figure~\ref{CICycle} illustrates the main elements and the data flow that composes DeepOrder. 
	
	First, historical datasets are collected out of existing CI cycles.  
	Next, in the training phase, feature extraction is applied to structure the datasets in a similar form, i.e., to contain similar features and label values (test case priorities).  
	
	Further, observing that many datasets are usually imbalanced~\cite{R32} (which is the case for all four real-world datasets investigated in this paper), we perform data augmentation, in order to strengthen model training. Data imbalance, in this case, relates to a low ratio of failed test executions compared to passed test executions. Given that historical failures can indicate a relevant test case (in the context of test prioritization) better than passes~\cite{R3}, we augment such imbalanced datasets to improve the problem of insufficient representation of the most relevant test cases for test prioritization. The goal is to change the target variable distribution to force the learning algorithm to focus on the rare and interesting cases, which in our scenario are the failure test scenarios. Without using such synthetic data, the loss function at the output layer would average overall samples and hence would lead towards better learning for well-represented regions in the dataset, and poor learning of the under-represented values. By creating additional synthetic data to complement existing industrial datasets (see section \ref{sec:synthetic}), we largely improved model training in DeepOrder. Next, a deep learning model is trained to prioritize test cases until the Mean Square Error (MSE) is less than $0.0001$. 
	
	Finally, once trained, the deep learning model can be deployed in CI testing to perform the regression test prioritization of previously unseen test suites. Feature extraction 
	is applied over the unseen test suite and DeepOrder computes the priorities for test cases, which are used for test case ranking depending on the available time budget.
	
	
	\subsection{DeepOrder Model Architecture}~\label{sec:Architecture}
	DeepOrder involves a multi-layer neural network~\cite{R12} with a small number of hidden layers and a reasonable number of neurons, so that its training does not require extensive time and computing resources~\cite{R13}. The goal is to handle the non-linearity of the data and all the features of test cases for voluminous dataset, without incurring high time-complexity of training or test prioritization. 
	
\begin{figure*}[!bthp]
		\centering
		\includegraphics[width=5.7in, keepaspectratio]{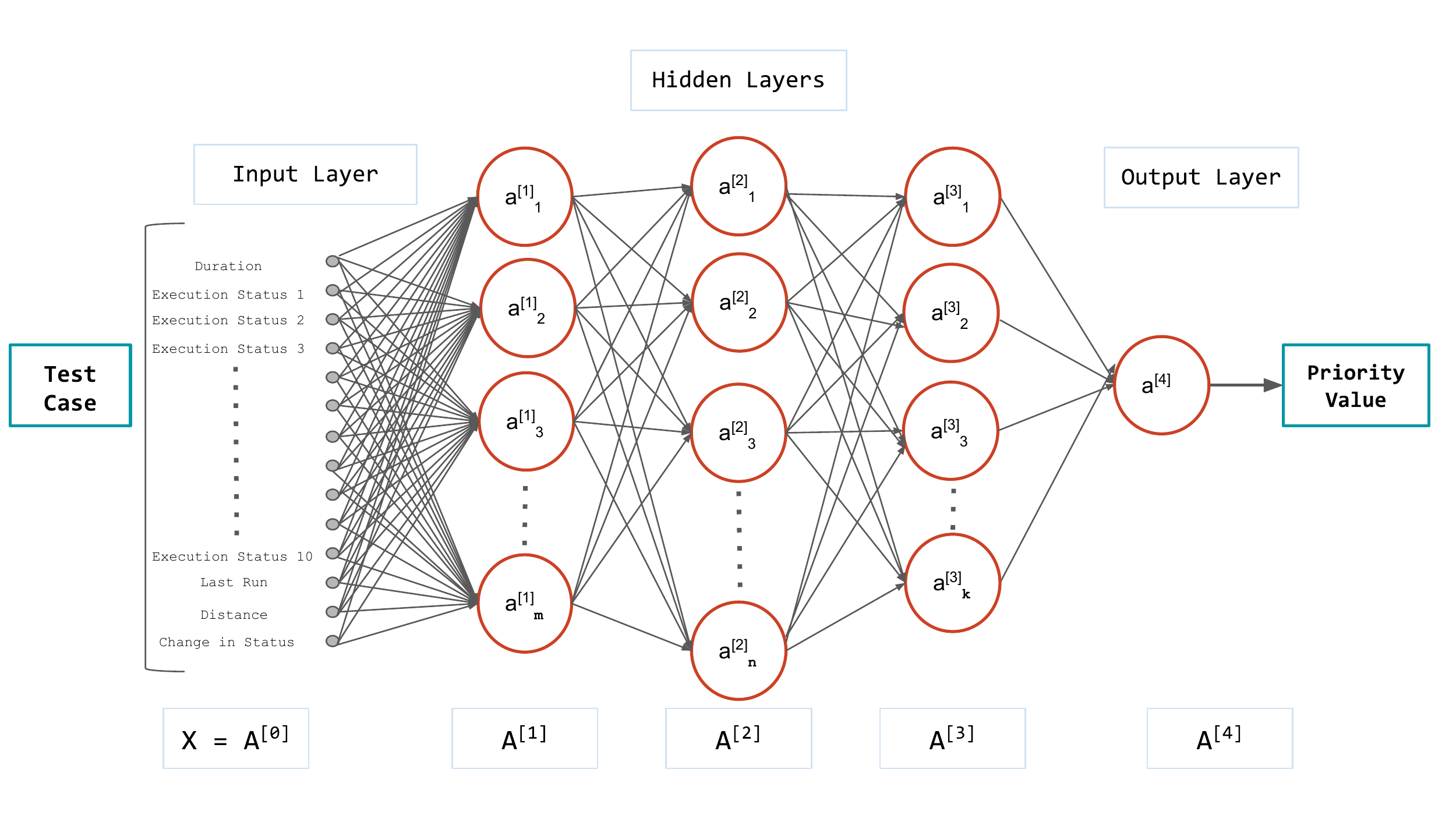}
		\caption{DeepOrder Model Architecture: Features defining a test case are passed through the input layer in DeepOrder to predict the test case priority value in the output layer. The number of hidden layers and the exact number of neurons in the input and output layers have been determined experimentally to cope with time and resource constraints.}
		\label{DeepOrder}
	\end{figure*}

	The DeepOrder architecture is shown in Figure~\ref{DeepOrder}. 
	Features in the input layer represent test cases and their properties including test execution time, previous execution status in a given number of CI cycles, time of the last execution. The output layer generates a real value, which is the computed priority for each test case. Since it is a regression model, the loss is calculated based on each iteration such that the weights within the model are updated. This process is repeated until we reach the point where the model output is close to the actual labels. 
	In the following, we present various elements of the DeepOrder architecture.

	\subsubsection{Input Layer}
	The input layer is defined by test case features, which feed the model for its training and execution. For each test case $t_i$, DeepOrder uses 14 input features in training. The features include the test case \textit{ExecutionStatus} $ES_{(i,1)},\ldots,ES_{(i,10)}$, and four additional features: {\it Duration}, {\it LastRun}, {\it Distance}, and {\it ChangeInStatus}, which further describe test cases. 
	\textit{Duration} $d(t_i)$ is an average test case execution time computed across all its previous executions. The value in seconds is normalized down to a range of $(0,1)$ for better regression training. \textit{LastRun} denotes the absolute time when a test case was last run. The original format YYYY-MM-DD is transformed into a numerical floating-point number which can feed a DNN.
	%
	%
	%
	%
	\textit{Distance} 
	is an absolute difference of a test case execution status between the least recent and most recent CI cycle. 
	\textit{ChangeInStatus} 
	represents the number of times a test case execution status has changed from pass to fail in all its previous executions. 

	\subsubsection{Hidden Layers}
	We experimentally came up with an architecture consisting of $3$ hidden layers with $10, 20, 15$ neurons, after a large number of trials of different architectures. This architecture showed the best results for the various datasets later used in evaluation (see section \ref{sec:Evaluation}).
	Each hidden layer is computed using a linear bi-parameter function \( h(x)= \theta_{0} + \theta_{1}x \), where \textit{x} refers to the input from the previous layer, and $ \theta_0 $ and $ \theta_1 $ are the weight parameters. 
	
	\subsubsection{Activation Functions}
	DeepOrder uses Mish~\cite{R1} as an activation function in all hidden layers. Mish was recently designed to overcome the flaws of known activation functions such as ReLU~\cite{R23}. The function is defined as follows:
	\[ f(x) = x \times {\tt tanh} ({\tt softplus}(x)) = x \times {\tt tanh}(\ln (1 + e^{x})) \]
	where $x$ is multiplied by a transformation of $ x $, thus preserving small amount of negative information that helps to smooth information flow.
	
	\subsubsection{Hyperparameters}
	Hyperparameters in any deep learning method are selected and experimented through the training process to reach the highest model performance results. The DeepOrder model is trained using the hyperparameters shown in Table~\ref{tab:Hyperparameters}, as further detailed in Section~\ref{sec:TrainTest}.
	
	\subsubsection{Output Layer}
	Our target is to minimize the loss between the actual and predicted priority values for prioritizing test cases in the correct descending order. Only one node is used in DeepOrder's final layer since it is a regression-based model.
	
	\subsubsection{Backpropagation}
	Backpropagation is performed using Gradient descent~\cite{R14}. 
	DeepOrder uses Mean Square Error (MSE) to determine the loss of the network. 
	After the loss is calculated, we find the gradient with respect to each neuron's weight parameters using partial derivatives. 
	Using backpropagation we refine the weight parameters inside all the hidden layers of the DeepOrder network. The process is repeated until the regression model has converged entirely to the dataset.

	\section{Experimental Evaluation}~\label{sec:Evaluation}
	The experimental evaluation of DeepOrder is guided by the following research questions:
	
	\textbf{RQ1:} 
	How does DeepOrder perform in terms of fault-detection effectiveness? 
	
	\textbf{RQ2:} 
	How does DeepOrder perform in terms of time effectiveness?

	\textbf{RQ3:} Can increasing the length of test history used for test case prioritization increase fault-detection effectiveness?
	
	\textbf{RQ4:} What is the prediction performance of the DeepOrder model? Since it is a regression-based model, we evaluate the accuracy of the model as a linear regression model.
	
	We address the research questions in comparison with ROCKET, showing the improvement over the state-of-the-practice, and RETECS, showing the improvement over the state-of-the-art.
	
	\subsection{Experimental Setup}~\label{sec:Setup}
	To evaluate DeepOrder for the Cisco case study, we use the Cisco dataset. To improve the generalizability of the results, we use three additional industrial datasets from real-world CI environments and three synthetic datasets.
	
	\subsubsection{Industrial Datasets}~\label{sec:Dataset}
	The Cisco dataset 
	is a test suite used for testing video-conferencing systems, provided by Cisco Systems. 
	The other three industrial datasets are used for testing industrial robotics applications provided by ABB Robotics (Paint Control and Input/Output Control, noted as IOF/ROL) and the GSDTSR~\cite{R22} test suite, which is a dataset open-sourced by Google. 
	
	The details of the four industrial test suites are provided in Table~\ref{tab:Dataset}. For each test suite, we show the number of test cases, the total number of test executions, and the percentage of failed test executions. 
	The Cisco test suite has the smallest number of test executions, while the Paint Control and IOF/ROL test suites have more overall test executions, but fewer failed test executions relative to the overall number of test executions, as compared to Cisco. 
	The Google test suite is the largest of all with millions of test case executions, although with the lowest percentage of failed test executions. 

	\begin{table*}
		\begin{center}
			\caption{Overview of Industrial and Synthetic Datasets used in the training and evaluation of DeepOrder.} \label{tab:Dataset}
			\resizebox{!}{1.15cm}{
				\begin{tabular}{lllllll}
					\hline
					\multirow{2}{*}{Dataset} &
					\multicolumn{2}{c}{Test Cases} &
					\multicolumn{2}{c}{Test Executions} &
					\multicolumn{2}{c}{Failed Test Executions} \\
					& Industrial & Synthetic & Industrial & Synthetic & Industrial & Synthetic \\
					\hline
					Cisco & 320 &  327 &  1068 & 1341 & 0.43\% & 9\% \\
					
					Paint Control& 85 & 596  & 248150 & 340500 & 0.19\% & 2.6\% \\
					
					IOF/ROL& 1488 & 3596  & 149700 & 124420 & 0.28\% & 4.8\% \\
					
					GSDTSR & 5507  & - & 12439910 & - & 0.00250\% & -  \\
					\hline
				\end{tabular}
			}
		\end{center}
	\end{table*}

	\subsubsection{Synthetic Datasets (Data Augmentation)}~\label{sec:synthetic}
	As shown in Table~\ref{tab:Dataset}, the ratio of failed test executions is extremely low. To address the problem of insufficient representation of relevant test cases in the industrial datasets, we perform data augmentation. Specifically, we use SMOGN\cite{R31}, which is a technique for tackling imbalanced regression datasets by generating diverse new data points for the given data. SMOGN combines under- and over-sampling strategies to create well-balanced synthetic data over the original test suites. It uses traditional interpolation techniques, such as SMOTER \cite{R33} and SMOTER-GN (Gaussian Noise) \cite{R34}. More specifically, we can assume two bins, as $Bin_{failed}$ having failed test cases and $Bin_{passed}$ containing passed test cases. Due to the imbalanced nature of the datasets, $Bin_{passed}$ goes through random under-sampling, while $Bin_{failed}$ goes through oversampling techniques. The key idea is that given a set of nearest neighbors $k$ within $Bin_{failed}$, a seed example will be generated using SMOTER's interpolation if the distance in $k$ neighbors is present in a safe zone. Gaussian Noise, on the other hand, is used when the distance between the nearest-neighbor and seed example is above the threshold, leading towards an unsafe zone.
	
	By doing so, we generated and concatenated the additional datasets to the industrial datasets for the training of DeepOrder model. 
	We augmented Cisco, Paint Control, and IOF/ROl test suites. Due to the large size of the GSDTSR dataset, we excluded it from the process of synthetic data generation. Details of the synthetic datasets are provided in Table~\ref{tab:Dataset}.
	
	\subsubsection{Data Preprocessing}\label{sec:DataPre}
\begin{figure}[!bthp]
		\centering
		\includegraphics[trim=2cm 7cm 1.5cm 7cm, width=3.5in]{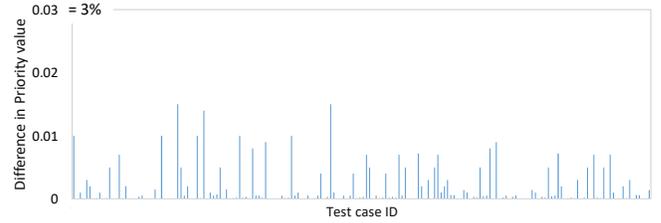}
		\caption{Absolute difference between $actual$ test case Priority values and Priority values $computed$ by ROCKET for the Cisco datasets.}
		\label{fig:difference}
	\end{figure}

	Data preprocessing was necessary to format all the four test suites in the same way, 
	for example, to make the number and type of features constant across all the files. 
	Features like \textit{Duration}, \textit{LastRun}, \textit{Execution Status} 
	are already provided in the test suites. \textit{Priority} value for each test case is available for the Cisco dataset. However, for the Google and ABB Robotics datasets, no historical test execution statuses are provided and thus test case \textit{Priority} values (which we need as labels in training) are not available. To obtain the labels for the latter datasets, we used a deterministic approach for test case prioritization ROCKET. To make sure that ROCKET is an appropriate approach for this purpose, we evaluated its performance to accurately compute priorities for the Cisco dataset (for which we have actual values available to compare against). As Figure~\ref{fig:difference} shows, ROCKET reliably computes test case \textit{Priority} values, with an average difference of 0.5\% and the maximum difference of 1.5\% between the actual and computed values.       
	
	While ROCKET is able to accurately compute test case priorities based on historical test execution data, the limitation of the approach is that it becomes computationally expensive for larger test suites, containing historical test data from longer test history and with a larger number of test executions. In particular, for the Google dataset, which has more than 12 million test case executions, ROCKET requires almost twice the amount of runtime compared to DeepOrder (see Figure~\ref{fig:totalrunningtime}-left). Furthermore, ROCKET requires approximately 70 times the amount of time required by DeepOrder for the task of prioritizing the Google dataset (see Figure~\ref{fig:totalrunningtime}-right). This circumstance is the reason we propose DeepOrder, to be able to more efficiently process large datasets and use longer test history, which we experimentally show to improve the fault-detection effectiveness of prioritized test suites (see section~\ref{sec:RQ4} and Figure~\ref{Effect}).   
	
	\subsubsection{Training and Testing Data}~\label{sec:TrainTest}
	DeepOrder model training is performed using both industrial and synthetically generated datasets. For the industrial datasets, we applied a time-based split, where for a given point in time $t$, we used data before $t$ to train the model and make predictions about test case priorities for CI runs after $t$. The synthetically generated datasets were split between train and validation sets using the 80/20 split values. 
	The testing of the DeepOrder model is performed only on the industrial test suites. 

	\textbf{Hyperparameter tuning:} The performance of our deep learning model depends on different parameters, including the number of hidden layers and their sizes, and the number of epochs. Therefore, we had to carefully tune the parameters to develop a well performing model. Table~\ref{tab:Hyperparameters} shows the hyperparameters of the final experimented version of the DeepOrder model. The model and its parameters 
	are constant throughout all the datasets. The values of the weight parameters that predict test case priorities are achieved with the given settings after several experiments.
	
	\begin{table}[htbp]
		\caption{Hyperparameters for the DeepOrder model.} \label{tab:Hyperparameters} 
		\begin{center}
			\resizebox{!}{2.3cm}{
				\begin{tabular}{lll}
					
					\toprule
					
					\textbf{Stage} & \textbf{Hyperparameter} & \textbf{Value} \\ \midrule
					
					Initialization & Weights & Xavier (Default)~\cite{R25} \\ 
					
					Activation Function & Mish~\cite{R1} & Default \\ 
					Hidden Layer 1 & - & 10 neurons \\ 
					Hidden Layer 2 & - & 20 neurons \\ 				
					Hidden Layer 3 & - & 15 neurons \\ 				
					\midrule
					\multirow{4}{*}{Training} & Epochs & 1000 \\  
					
					& Loss Function & Mean Sqaured Error~\cite{R10} \\ 
					
					& Optimizer & Adam~\cite{R24} \\ 
					& Learning Rate & 0.001 \\ 
					\midrule
					\multirow{3}{*}{Model Parameters} & Total Parameters & 631 \\ 
					
					& Trainable Parameters  & 631 \\ 
					
					& Non-Trainable Parameters & 0  \\ \bottomrule
					
				\end{tabular}

			}
			
		\end{center}
	\end{table}

	\subsubsection{Code Implementation}~\label{sec:code}
	DeepOrder is implemented using Python~\cite{R6}. The deep neural regression model is written on top of Keras~\cite{R2}, along with Tensorflow~\cite{R4} running in the background for GPU support. The DeepOrder code implementation, test sets and experimental results are available on Github\footnote{Repository: https://github.com/AizazSharif/DeepOrder-ICSME21}. 

	
	\subsection{Evaluation Metrics}~\label{sec:Metric}
	For the purpose of evaluating the fault detection effectiveness of DeepOrder and comparing it with 
	previous research work, we used APFD~\cite{R8} and NAPFD~\cite{R9} metrics. 

	%
	Next, since DeepOrder is a regression model, one of the most common metrics to evaluate the prediction accuracy of the model is Mean Squared Error (MSE)~\cite{R10}. 
	
	\begin{table}[!htbp]
		\centering
		\caption{Metrics for evaluationg DeepOrder.} \label{tab:Metrics}
		\begin{center}
			{
				\begin{tabular}{ll}
					\toprule
					\multicolumn{1}{l}{\textbf{Metric}} &   \multicolumn{1}{l}{\textbf{Description}}   \\ \midrule	
					\multicolumn{1}{l}{APFD} & \multicolumn{1}{l}{Average Percentage of Faults Detected} \\ 	
					\multicolumn{1}{l}{NAPFD} &  \multicolumn{1}{l}{Normalized Average Percentage of Faults Detected} \\	\midrule
					\multicolumn{1}{l}{MSE} &  \multicolumn{1}{l}{Mean Squared Error} \\	\midrule
					\multicolumn{1}{l}{FT} & \multicolumn{1}{l}{Time to detect the first fault} \\ 	
					\multicolumn{1}{l}{LT} &  \multicolumn{1}{l}{Time to detect the last fault} \\	
					\multicolumn{1}{l}{AT} &   \multicolumn{1}{l}{Average time spent to detect all faults in regressing testing} \\ 	
					\multicolumn{1}{l}{PT} &   \multicolumn{1}{l}{Time to process a dataset, train a model, and validate results} \\
					\multicolumn{1}{l}{RT} &   \multicolumn{1}{l}{Time to prioritize a test suite} \\
					\multicolumn{1}{l}{TT} &   \multicolumn{1}{l}{Total algorithm running time} \\ 
					\bottomrule				
			\end{tabular}}
		\end{center}
	\end{table}
	Finally, since time plays an important factor in the CI process~\cite{R21}, and since, with DeepOrder, we aim to increase fault detection in the early stages of testing, we use 6 other metrics shown in Table~\ref{tab:Metrics} to evaluate the time-effectiveness of DeepOrder.
	
	\section{Results and Analysis}\label{sec:Results}
	In this section, we present the experimental results answering the 4 research questions. We also analyse the threats to validity of our results.
	
	\subsection{RQ1: Fault detection effectiveness}~\label{sec:RQ1}
\begin{figure}[!bthp]
		\centering
		\includegraphics[width=2.7in, keepaspectratio]{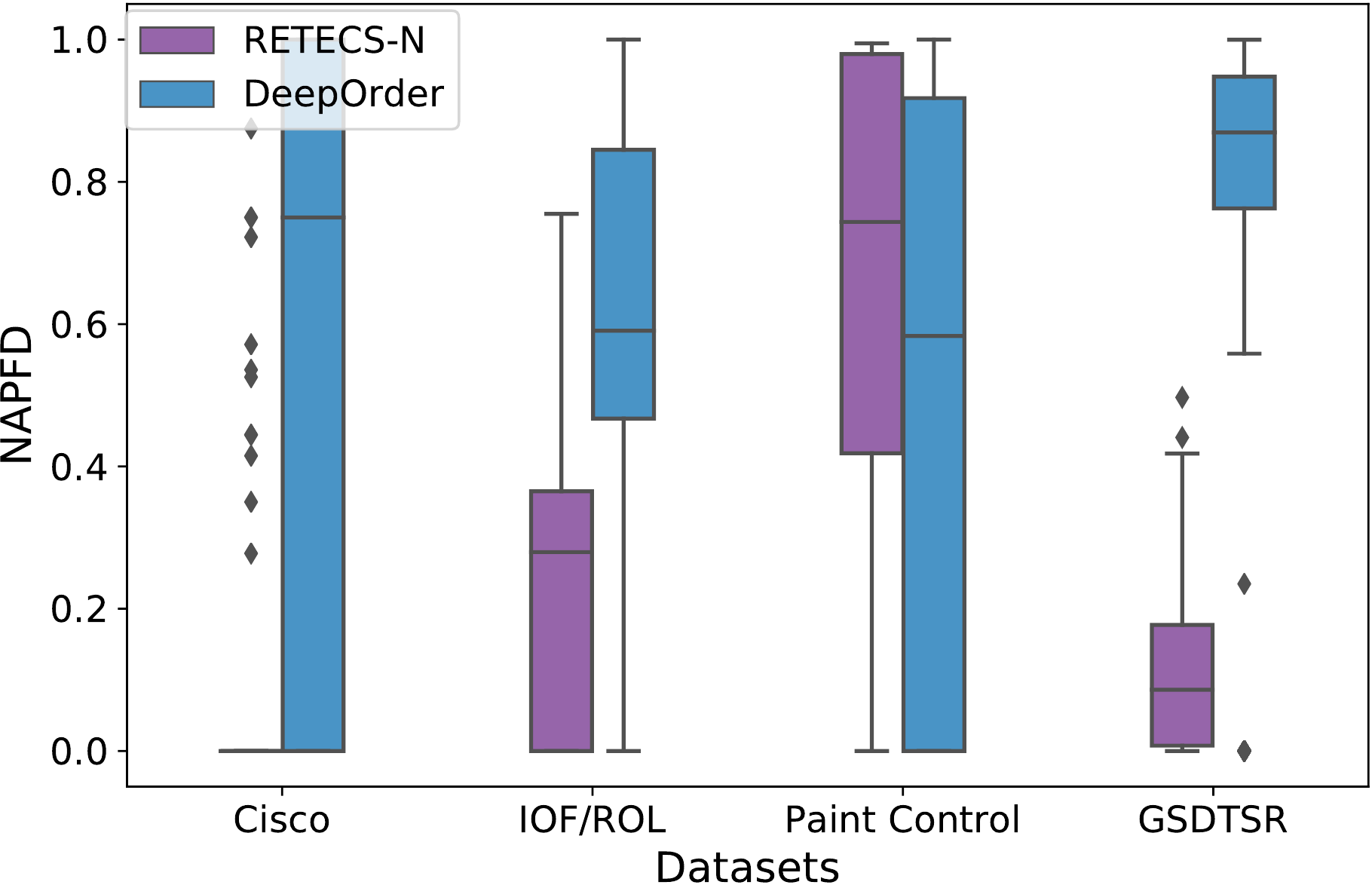}
		\caption{Comparison of DeepOrder and RETECS-N in terms of NAPFD across 4 industrial datasets.}
		\label{fig:boxplot}
	\end{figure}

	We evaluate the fault detection effectiveness of DeepOrder in comparison with the SoA approach RETECS. Figure~\ref{fig:boxplot} 
	shows the model performance in terms of NAPFD. 
	For proper evaluation, we used the RETECS code to produce the results on 
	4 industrial datasets. 
	There are two configurations of RETECS available, one using network-based test case failure reward (RETECS-N) and another using tableau-based time-ranked reward (RETECS-T). RETECS-N was reported to outperform RETECS-T~\cite{R3} and therefore we compare DeepOrder with RETECS-N. Figure~\ref{fig:boxplot} shows NAPFD for DeepOrder in comparison with RETECS-N. As we can see, RETECS-N only performs better on Paint Control. However, we can observe better overall performance of DeepOrder compared to RETECS-N in terms of fault detection effectiveness. DeepOrder performs consistently across all 4 datasets and avoids NAPFD outliers. 
	
	For the purpose of comparison with other approaches for RTP in the future, Table~\ref{tab:Results} shows the average APFD for DeepOrder on 4 industrial datasets.

	\subsection{RQ2: Time effectiveness}~\label{sec:RQ2}
\begin{figure}[!bthp]
		\centering
		\includegraphics[width=3.3in, keepaspectratio]{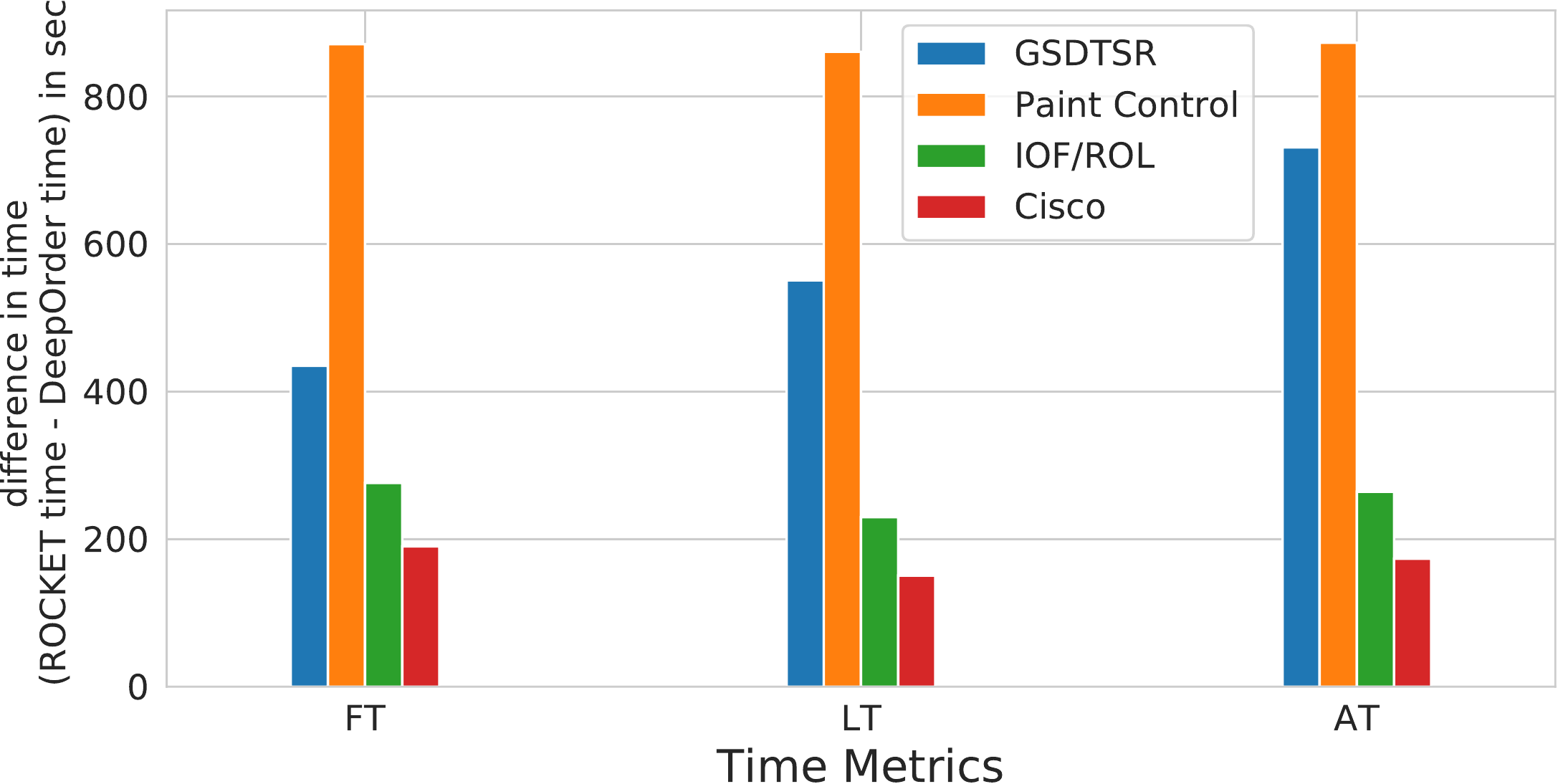}
		\caption{Comparison of DeepOrder and ROCKET in terms of the difference in time based metrics across 4 industrial datasets. Positive value indicates better performance of DeepOrder.}
		\label{fig:FTLTAT}
	\end{figure}

	To answer RQ2, first we compare DeepOrder with ROCKET in terms of FT, LT, and AT metrics. The values of the individual metrics for DeepOrder are presented in Table~\ref{tab:Time}, and the comparison with ROCKET is shown in Figure~\ref{fig:FTLTAT}. Y-axis represents the difference in time between ROCKET and DeepOrder ($ROCKET\_time$ - $DeepOrder\_time$), for FT, LT, and AT metrics. The positive values of the difference show that DeepOrder outperforms ROCKET in terms of time-effectiveness, for all four datasets, answering RQ2. 
	
	\begin{table}[!htbp]
		\centering
		\caption{Time-based metrics for DeepOrder.} \label{tab:Time}
		\begin{center}
			\resizebox{!}{0.95cm}{
				\begin{tabular}{lllllll}
					\toprule
					\multicolumn{1}{l}{Dataset} &   \multicolumn{1}{r}{\textbf{FT [s]}} & \multicolumn{1}{r}{\textbf{LT [s]}} & \multicolumn{1}{r}{\textbf{AT [s]}} & \multicolumn{1}{r}{\textbf{PT [s]}} & \multicolumn{1}{r}{\textbf{RT [s]}} & \multicolumn{1}{r}{\textbf{TT [s]}} \\ \midrule
					
					\multicolumn{1}{l}{Cisco} & \multicolumn{1}{r}{1.493} & \multicolumn{1}{r}{61.771} & \multicolumn{1}{r}{31.632} & \multicolumn{1}{r}{7} & \multicolumn{1}{r}{0.009} & \multicolumn{1}{r}{19}\\ 
					
					\multicolumn{1}{l}{Paint Control} &  \multicolumn{1}{r}{2.766} & \multicolumn{1}{r}{14.345} & \multicolumn{1}{r}{1.390} & \multicolumn{1}{r}{3} & \multicolumn{1}{r}{0.227} & \multicolumn{1}{r}{31}\\ 
					
					\multicolumn{1}{l}{IOF/ROL} &   \multicolumn{1}{r}{3.021} & \multicolumn{1}{r}{22.427} & \multicolumn{1}{r}{1.522} & \multicolumn{1}{r}{14} & \multicolumn{1}{r}{0.263} & \multicolumn{1}{r}{38}\\ 								
					
					\multicolumn{1}{l}{GSDTSR} &   \multicolumn{1}{r}{120} & \multicolumn{1}{r}{473.489} & \multicolumn{1}{r}{58.651} & \multicolumn{1}{r}{39} & \multicolumn{1}{r}{13} & \multicolumn{1}{r}{716}\\ 
					\bottomrule				
			\end{tabular}}
		\end{center}
	\end{table}
	
\begin{figure}[!bthp]
		\centering
		\includegraphics[trim=0cm 0cm 0cm 0cm, width=3.3in]{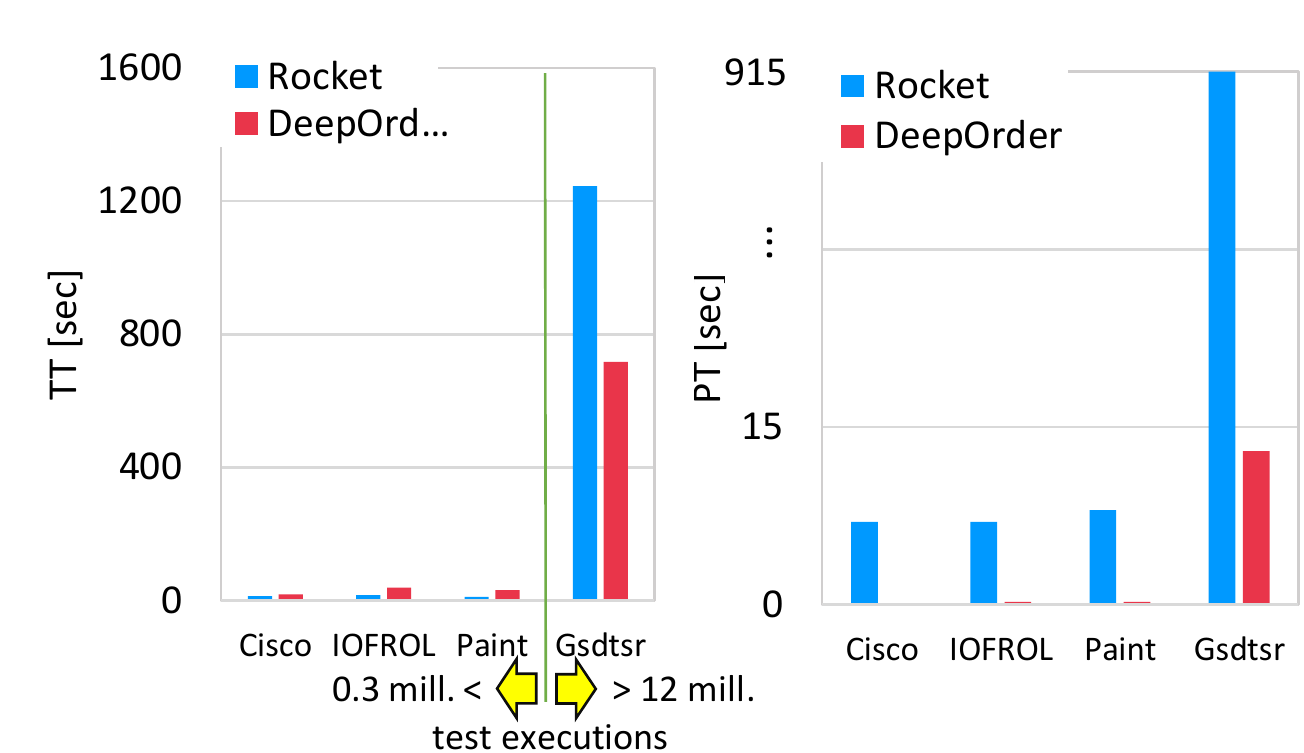}
		\caption{Comparison of TT and RT metrics for ROCKET and DeepOrder on 4 industrial datasets.}
		\label{fig:totalrunningtime}
	\end{figure}

	Next, we compare DeepOrder with ROCKET in terms of the TT and RT metrics. Regarding the TT metric, as can be seen in Figure \ref{fig:totalrunningtime}, when we are dealing with the datasets containing a large number of test executions (e.g. 12 million in the case of the Google dataset) DeepOrder is superior to ROCKET, requiring almost half the time needed by ROCKET. This result confirms our starting hypothesis that DeepOrder is an efficient approach to deal with voluminous test history in test prioritization. Regarding the RT metric, we can see in Figure \ref{fig:totalrunningtime} that DeepOrder outperforms ROCKET for all four datasets, regardless of the number of test executions. From these results, we can conclude that if we are dealing with test suites containing a small number of executions and have the requirement for retraining a learning model for RTP in every CI cycle, DeepOrder is not the most time-efficient approach. However, if we are dealing with test suites containing a large number of executions, which is often the case for real-world CI environments, or when we do not need to retrain a learning model for RTP in every CI cycle, DeepOrder is a time-efficient approach. Most models are retrained only periodically, when there start to be deviations of the distributions of the input features from the distributions of the training data. The frequency of model retraining depends on the changes in the software under test, i.e. how often software features change, which is not expected to happen in every CI cycle. Besides, model retraining is less time demanding than the original model training, as it usually involves using a new training set, while keeping the original model architecture and hyperparameters.

	Finally, we compare DeepOrder with RETECS in terms of the TT metric, on 4 industrial datasets.
	The total running time of the DeepOrder algorithm on all 4 datasets was approximately \textbf{0.5 hours}, while RETECS required around \textbf{25 hours}. RT time for RETECS was 20 times higher than for DeepOrder in the base case scenario.

	\subsection{RQ3: Effect of longer test history}~\label{sec:RQ4}
	The existing SoA machine learning based approach for RTP~\cite{R3} performs the best when using test history consisting of the last four cycles. In contrast, DeepOrder can use test history from any number of test cycles for test case prioritization. 
	We evaluate the ability of DeepOrder to increase the fault detection effectiveness of prioritized test suites as a result of using longer test history. 
	
\begin{figure}[!bthp]
		\centering
		\includegraphics[width=3.3 in, keepaspectratio]{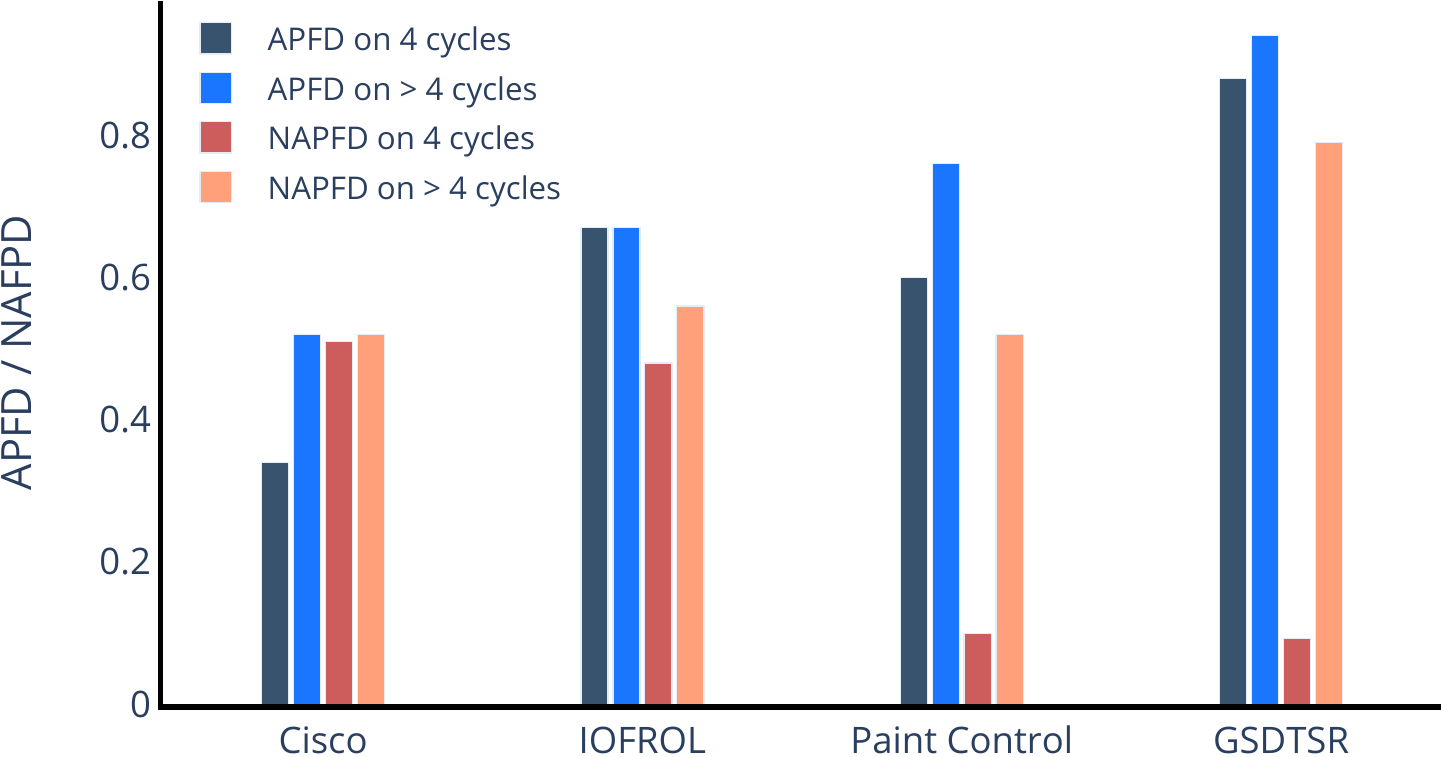}
		\caption{Effect of test history length on fault detection effectiveness. Using longer history for test case prioritization increases fault detection effectiveness in terms of APFD and NAPFD metrics.}
		\label{Effect}
	\end{figure}

	Figure~\ref{Effect} compares NAPFD and APFD for two test suites prioritized by DeepOrder: one using the test history of exactly 4 last test cycles and another using the test history of longer than the 4 last test cycles. We can see from the figure that increasing the length of test history used in prioritization increases the effectiveness of test case prioritization in terms of NAPFD and APFD, for all four datasets, answering RQ3.  
	
	These findings are contradicting to the observations made by the authors of RETECS, who suggested that longer history does not necessarily correspond to better performance. We explain these findings by the fact that DeepOrder and RETECS are architecturally different learning algorithms. For RETECS, longer history increases the state space of possible test case representations, which further increases the complexity and requires more information to be available to the reinforcement learning agent, who has to process earlier execution results differently than more
	recent ones. On the other hand, DeepOrder uses a deep learning model, which in nature increases prediction performance when more data is available.  
	
	\subsection{RQ4: Prediction accuracy}~\label{sec:RQ5}
	First, we evaluate the prediction accuracy of the DeepOrder model in terms of MSE. Table~\ref{tab:Results} displays MSE for the DeepOrder regression model on the 4 industrial datasets. The results show that DeepOrder achieves low MSE values for a deep regression model, which indicates a good model performance. DeepOrder also reaches high $R^2$, suggesting a higher model fitness over the dataset. Furthermore, since DeepOrder learns a regression line over the set of test cases, the values of the standard deviation clearly show that the majority of the test cases lie near the regression line learned by the deep neural network.
	
\begin{figure}[!bthp]
		\centering
		\includegraphics[width=3.7in, keepaspectratio]{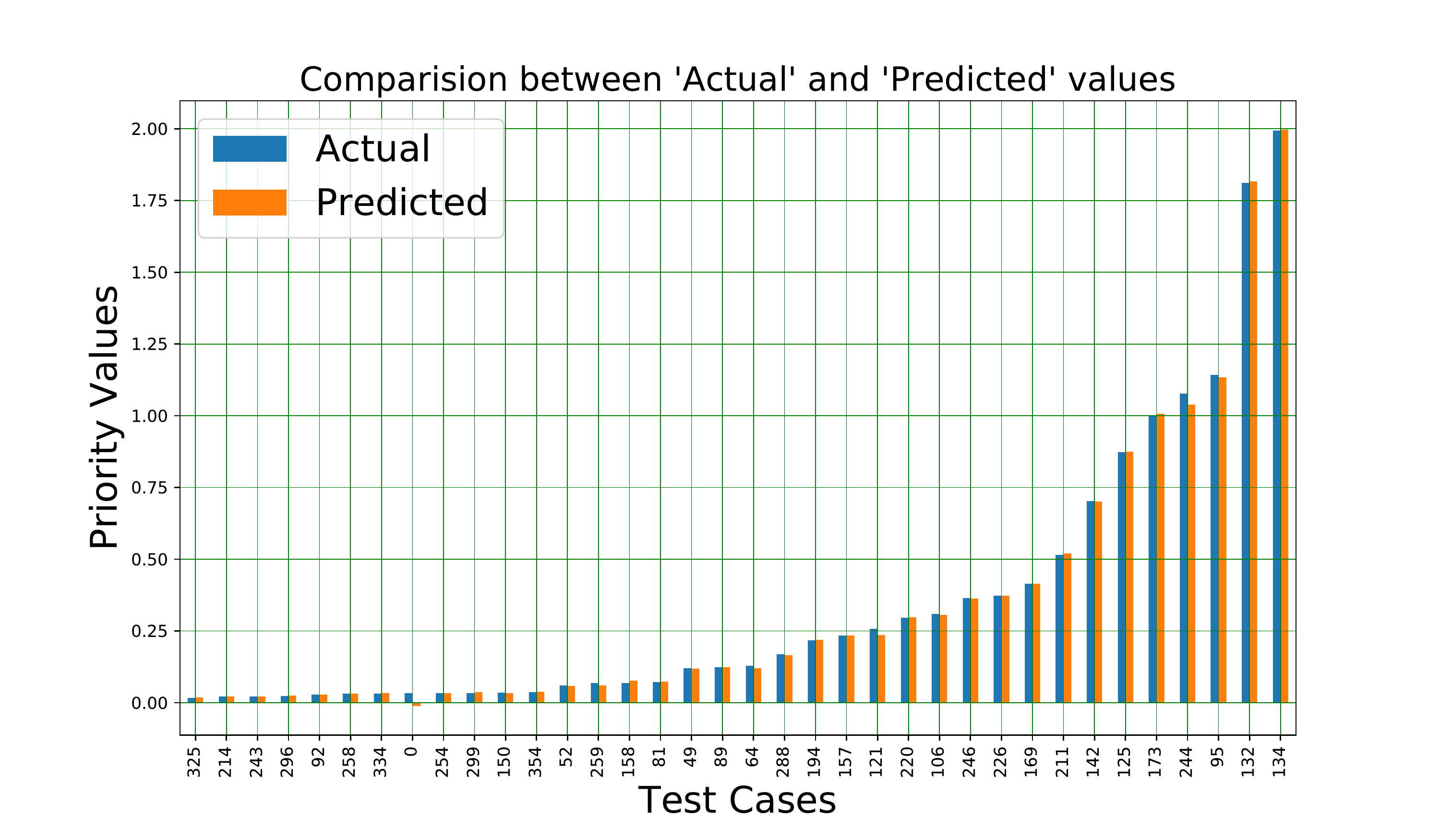}
		\caption{Prediction performance of DeepOrder in terms of $actual$ vs $predicted$ values for the IOF/ROL dataset.}
		\label{PriorityyPlot}
	\end{figure}

	Second, we compared the test case priority values that were predicted by DeepOrder against the ground truth priority values. Because IOF/ROL, Paint Control, and Google dataset do not contain historical test case execution statuses, we could not compare against the ground truth values coming from these industrial datasets. However, we compared the DeepOrder prediction accuracy against the \textit{Priority} values computed by ROCKET, as we have previously shown that ROCKET computes \textit{Priority} values that are very close to the original ground truth values (see Figure \ref{fig:difference}). 
	Figure~\ref{PriorityyPlot} shows the prediction accuracy of the DeepOrder model for the IOF/ROL dataset in a comparison with the surrogate ground truth values provided by ROCKET. The results indicate a good prediction performance of DeepOrder. 
	
\begin{figure}
		\centering
		\includegraphics[width=3.0in, keepaspectratio]{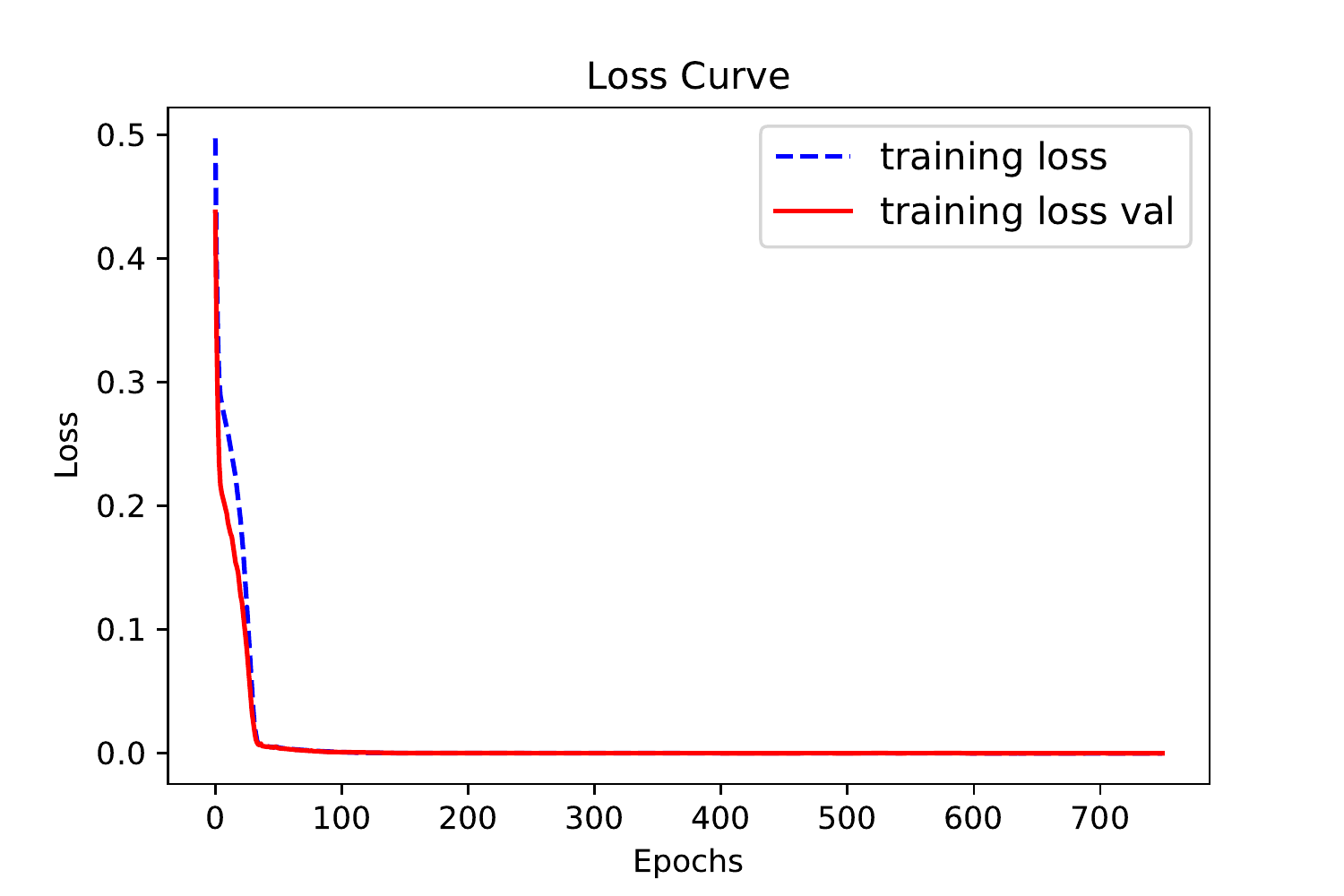}
		\caption{Loss function of DeepOrder on the Cisco dataset.}
		\label{Loss}
	\end{figure}

	Third, Figure~\ref{Loss} depicts the loss function of the DeepOrder model 
	on the Cisco dataset. The loss function represents the difference between the actual and expected output of the regression model and our goal is to minimize the loss function as low as possible. 
	The results in Figure~\ref{Loss} show that DeepOrder on the Cisco dataset converges towards the expected minimum loss values.
	
	In summary, the results shown in Table~\ref{tab:Results}, Figure~\ref{PriorityyPlot}, and Figure~\ref{Loss} indicate that DeepOrder model has a good prediction accuracy, answering RQ4.

	\begin{table}[!htbp]
		\caption{Prediction accuracy for DeepOrder on the 4 industrial datasets.} \label{tab:Results}
		\begin{center}
			\resizebox{!}{1.0cm}{
				\begin{tabular}{llllll}
					
					\toprule
					\multicolumn{1}{l}{Dataset} & \multicolumn{1}{r}{\textbf{MSE}} & \multicolumn{1}{r}{\textbf{$R^2$}} & \multicolumn{1}{r|}{\textbf{Std. Deviation}} &
					\multicolumn{1}{r}{\textbf{NAPFD}} & \multicolumn{1}{r}{\textbf{APFD}}  \\ \midrule
					
					\multicolumn{1}{l}{Cisco} & \multicolumn{1}{r}{0.000038} & \multicolumn{1}{r}{0.75} & \multicolumn{1}{r|}{0.014} &
					\multicolumn{1}{r}{0.52} & \multicolumn{1}{r}{0.52} \\ 
					
					\multicolumn{1}{l}{Paint Control} & \multicolumn{1}{r}{0.000123} & \multicolumn{1}{r}{0.82} & \multicolumn{1}{r|}{0.0079} & 
					\multicolumn{1}{r}{0.52} & \multicolumn{1}{r}{0.76}\\ 
					
					\multicolumn{1}{l}{IOF/ROL} & \multicolumn{1}{r}{0.000001} & \multicolumn{1}{r}{0.80} & \multicolumn{1}{r|}{0.006} & 
					\multicolumn{1}{r}{0.56} & \multicolumn{1}{r}{0.67}\\ 
					
					\multicolumn{1}{l}{GSDTSR} &  \multicolumn{1}{r}{0.000031} & \multicolumn{1}{r}{0.85} & \multicolumn{1}{r|}{0.00139} & 			 \multicolumn{1}{r}{0.79} & \multicolumn{1}{r}{0.94}\\ 
					\bottomrule

			\end{tabular}}
		\end{center}
	\end{table}

	\subsection{Threats to Validity}
	We identify the following threats to the validity of our experimental results.
	
	\textbf{Internal}: First, we might have introduced errors in our implementation of the DeepOrder algorithm. To mitigate this threat, we used a widely-used framework Keras for the algorithm development and we carefully examined the implementation. Furthermore, the DeepOrder implementation 
	is available for inspection and use on Github. Second threat to validity is the quality of the datasets used in DeepOrder training. For example, imbalanced datasets can lead to bias in model learning. To address this threat, we use additional datasets generated with data augmentation, using a technique for creating regression-based datasets. Third, a hyperparameter tuning plays a great role in achieving a good model performance. Using different hyperparameters may lead to different results. However, we experimented with many different configurations while training DeepOrder, and selected the best-performing set of hyperparameters.
	
	\textbf{External}: We use four industrial datasets to test DeepOrder, which limits the generalizability of the results. One out of these four datasets is provided by industry and three are publically available. Evaluating DeepOder on more industrial CI datasets is planned for future work. 
	
	\textbf{Construct}: We consider relevant for RTP only test cases that have a fail or pass status. Test cases that are not executed could also be considered as candidates for being prioritized higher. Furthermore, we consider relevant 14 features defining a test case. In different environments and case studies, a different set of features could be relevant. Both these threats will be considered in the DeepOrder extension.

	\section{Related Work}~\label{sec:RelatedWork}
	Extensive amount of work has been reported addressing the RTP problem. We broadly categorize these approaches into two groups: machine learning (ML)-based and non-machine learning-based. 
	
	\textbf{ML-based}: ML nowadays is often used in software engineering for test automation. 
	In~\cite{R15}, Machalica et al. propose a boosted decision tree method for test selection in CI environment. Their proposed solution is to learn a classifier based on a changed code and a subset of test cases, which predicts the probability of a test case failing. 
	The experiments are performed on Facebook's CI system. 
	Another work ~\cite{R16} uses SVM as a machine learning binary classifier to select and rank tests in the industrial study of Salesforce. 
	In contrast to DeepOrder, this work does not take the time factor into consideration, which is important for RTP in CI.
	
	In~\cite{R20}, the authors use regression learning classification for predicting how much branch coverage will be made in test execution (before tests are executed). 
	The authors use Huber regression, SVM, and multilayer perceptron for training their regression models. Another work reported in \cite{R21} proposes a predictive test prioritization based on XGBoost. 
	It applies distribution analysis of test cases to compare their results with the fault detection ability of actual regression testing. Their approach is integrated and tested as a practical case study. They have mixed results when evaluated on industrial test suites. 
	The work in \cite{Mar} uses regression trees to minimize test case redundancy in CI testing and improve the cost-effectiveness of CI. Very recent work in~\cite{R27} proposes machine learning for the generation and prioritization of test cases for a user interface. Their approach concludes that K-Nearest Neighbors performed well among the rest of the machine learning classification models on their collected historical data from user interface design. 
	
	The authors in~\cite{R3} propose a reinforcement learning approach for RTP in CI, called RETECS. They use industrial datasets from Google and ABB Robotics to develop three reward function-based models. 
	As shown in our experiments, DeepOrder outperforms RETECS in both fault detection-effectiveness and time-effectiveness.  
	A recent work in \cite{Bertolino} reports a comprehensive study comparing  learning-to-rank and ranking-to-learn (RTL) algorithms. The authors apply a classification based on 4 classes and use a class level dependency graph, which is a different form of dataset than ours, and thus not applicable in our case. This study found that reinforcement learning based algorithms in the RTL approach they investigate take much longer than the rest of the ML algorithms, which is in line with what we have observed with RETECS. Another method proposed in~\cite{R29} introduces COLEMAN, an approach based on Multi-Armed Bandit (MAB), which is considered to be very close to reinforcement learning, and thus susceptible to high time complexity.
	
	\textbf{Non-ML-based}: 
	In~\cite{R18} researchers use Ruby on Rails datasets from Travis CI and Google's GSDTSR dataset to perform test case prioritization. They use commits instead of test cases in a CI environment, so that the commits that are more important can be executed for future code commits. Another approach reported in~\cite{R19} works on understanding and improving modules and class level test suite executions to create a hybrid method for regression test selection. The hybrid technique is then evaluated in detail with the module and class-level techniques to see the ratio of fault detection. 
	In \cite{Kim:2002pi} authors propose a technique that uses historical information from test cycles to select test cases that must be executed for a new version of software. However, the technique does not compute the order of test cases, which is necessary for the RTP problem. 
	The authors in \cite{Noor2015jt} define a set of metrics that estimate test cases quality using their similarity to the previously failing test cases. The similarity is expressed in terms of code coverage, which makes this approach inapplicable for our datasets, as code is unavailable. 
	In \cite{Mirarab}, the authors propose the the combination of Integer Linear Programming and greedy methods for multi-criteria test case prioritiztion. 
	In~\cite{R28} the authors propose an approach for multi-perspective regression test ranking by extending the work of~\cite{R5}~\cite{7928010}~\cite{7816510}. The approach is meant to optimize the regression test selection using a business, performance, and technical perspective for a faster fault detection. The work in \cite{8377636} proposes to combine fault-based and risk-based test prioritization to improve time-effectiveness of CI testing.
	
	While many approaches have been proposed for RTP, many of them are computationally complex or difficult to apply in practice, as they require different test parameters. DeepOrder, by contrast, requires only historical results of test case executions in previous CI cycles. Furthermore, in order to fairly compare with these proposed approaches, we would need to have their source code implementation, which is not available for many of them.

	\section{Conclusion \& Discussion}~\label{sec:Extensions}
	In this work, we present a regression based model for RTP, which uses a deep learning architecture at its core. The model learns to predict the priority of test cases in a given test suite using various features, including a test case duration and historical execution status. The model is evaluated using 4+3 datasets, showing a better performance compared to the industry practice and SoA approaches, with respect to fault-detection effectiveness and time-effectiveness. 
	
	\textbf{Model selection}: DeepOrder uses a deep learning model, while there could be other machine learning models suitable for the RTP problem. Our motivation for choosing a deep learning neural network with a small number of hidden layers and a reasonable number of neurons was to incur low training complexity, but still provide the flexibility to efficiently deal with large-scale datasets, such as the Google dataset used in the experiments. 
	Our deep learning model has shown to avoid the time-complexity incurred by reinforcement learning models for RTP (see experimental results).
	
	\textbf{Future work}: One aspect that can be further explored in DeepOrder is leveraging more features of test cases or additional information like changes in source code, version control, and code coverage. With such data, our model could learn the relationship between test cases and the code modules on which the tests were  executed. Another aspect that can be further explored is accounting for  
	flaky tests. Furthermore, we could take into consideration non-executed test cases during test case prioritization. 
	Yet another aspect of future work could be evaluating DeepOrder on new industrial CI datasets, improving the generalizability of our results. Furthermore, DeepOrder can be evaluated by deploying the algorithm in a real CI testing to compare the results with previous RTP techniques. In the CI environment, continuous change in code and unseen test cases can be used to evaluate how long DeepOrder can perform well before it needs to relearn on new test cases. This would turn the algorithm into the state of online learning.
	
	\section{Acknowledgment}
	This work is supported by the Research Council of Norway through T3AS project.

	\bibliographystyle{IEEEtran}
	\bibliography{sample-base}

\end{document}